\documentclass[sigplan,screen]{acmart}

\AtBeginDocument{%
  \providecommand\BibTeX{{%
    \normalfont B\kern-0.5em{\scshape i\kern-0.25em b}\kern-0.8em\TeX}}}

\setcopyright{none}
\renewcommand\footnotetextcopyrightpermission[1]{} 

\usepackage{hyperref}
\usepackage{multirow}
\usepackage{makecell}

\begin{document}

\title{The use of Semantic Technologies in Computer Science Curriculum: A Systematic Review}


\author{Yixin Cheng}
\email{Yixin.Cheng@anu.edu.au}
\affiliation{%
  \institution{The Australian National University}
  \streetaddress{Street}
  \city{Canberra}
  \country{Australia}
  \postcode{Postcode}
}

\author{Bernardo Pereira Nunes}
\email{Bernardo.Nunes@anu.edu.au}
\affiliation{%
  \institution{The Australian National University}
  \streetaddress{Street}
  \city{Canberra}
  \country{Australia}}

%
\renewcommand{\shortauthors}{Cheng and Nunes}

%
\begin{abstract}
Semantic technologies are evolving and being applied in several research areas, including the education domain. This paper presents the outcomes of a systematic review carried out to provide an overview of the application of semantic technologies in the context of the Computer Science curriculum and discuss the limitations in this field whilst offering insights for future research. A total of 4,510 studies were reviewed, and 37 were analysed and reported. As a result, while semantic technologies have been increasingly used to develop Computer Science curricula, the alignment of ontologies and accurate curricula assessment appears to be the most significant limitations to the widespread adoption of such technologies.
 
\end{abstract}



\maketitle
\pagestyle{plain}
\section{Introduction}
The Semantic Web (SW) is a decentralised global information space for sharing machine-readable data with minimal interoperability and integration costs~\cite{Imran2016, Yu2007}. It aims at enabling machines to understand and answer the requests of people and machines through several standards and technologies\footnote{\url{https://www.w3.org/standards/semanticweb/}}. For instance, the Resource Description Framework (RDF) is the standard data model to represent information\footnote{\url{https://www.w3.org/TR/rdf-concepts/}}~\cite{Klyne2006} whereas the Web Ontology Language (OWL) is the standard language for defining vocabularies/ontologies on the Web~\cite{Allard}.

Ontologies --- \emph{a formal representation of the shared knowledge of a domain}~\cite{Piedra2018} --- are one of the core technologies of the Semantic Web. Analogous to the concept of ontologies is the concept of curriculum in the educational domain. Cox~\cite{Cox2005}, for example, defines a curriculum as a selection and organisation of knowledge for educational purposes used to represent the shared agreement on what learners should know in different stages of life. Another definition is provided by Chung and Kim in \cite{Chung2014} where they state that a curriculum is a course blueprint designed to guide students' learning~\cite{Chung2014} through a planned sequence of subjects~\cite{Nuntawong2016}. As a technology, ontologies can support curriculum design by representing domain knowledge for educational purposes. Cassel et al. \cite{10.1145/1345375.1345439}, for instance, proposed the ontology of computing to support the development of a Computer Science curriculum while Wang et al. \cite{Wang2019} proposed the computer course architectural ontology system.

This paper presents a systematic review on the use of semantic technologies in the context of Computer Science Education. We review existing initiatives using semantic technologies for curriculum design as well as the key challenges for their widespread adoption. The following research question guided the development of this work:

\begin{itemize}
    \item What are the semantic technologies used in the context of Computer Science curriculum? 
\end{itemize}

A total of 4,510 papers were retrieved from mainstream digital libraries and, after detailed analysis, 37 papers were found to be relevant. The reviewed articles focus on the use of ontologies and semantic technologies for curriculum design, curriculum interoperability and analysis. Here, we report on the tools/frameworks, datasets, vocabularies/ontologies, as well as the challenges of using and adopting semantic technologies in curriculum design in Computer Science to leverage new research opportunities.

The remainder of this paper is structured as follows: Section \ref{sec:related_work} reviews closely related surveys to semantic technologies in Education. Section \ref{sec:research_methodology} describes the research methodology used to carry out this systematic review. Section \ref{sec:overview} provides an overview of employing semantic technologies in Computer Science curriculum. Section \ref{sec:challenges} discusses the challenges and, finally, Section \ref{sec:conclusion} concludes the paper.

\section{Related Work}
\label{sec:related_work}

Krieger and R\"osner~\cite{Krieger2011} present a comprehensive review of the effects and applications of semantic technologies in e-learning. They discuss the construction of models with educational ontologies where they provide an ontology-wise overview in a Web-based educational environment. According to their work, the educational semantic web services include adaptive learning and e-assessment. Finally, they raise a few issues of adopting semantic technologies for data integration such as trust and credibility in data from multiple sources.

Another relevant survey was conducted by Pereira et al.~\cite{Crystiam2018}, where they reviewed various applications of Linked Data in the context of Education, including datasets, tools and common vocabularies/ontologies. They also identified the challenges in each stage of the process and provided an overview of the development in that area.

In another survey, Navarrete and Lujan-Mora~\cite{Navarrete2015} focused on the use of Linked Data to enrich Open Educational Resources (OER). Their study described how OER Web sites can adopt the Linked Data principles through the 5-star deployment scheme to enhance discovery, retrieval, reuse and remix of OERs.

Regarding ontology-related studies, Stancin et al.~\cite{Stancin2020} surveyed ontologies used in educational contexts. Their survey is a valuable contribution to curriculum modelling and management describing learning domains, educational data and e-learning services using ontologies. They also provide a clear definition of the term ontology along with traditional methodologies for their creation in the field of Education. Likewise, Tapia-Leon et al.~\cite{Tapia2018} surveyed the application of ontologies in Higher Education institutions and proposed guidelines for ontology creation.

Despite the contributions presented in previous studies, semantic technologies in Curriculum Design, specifically applied to Computer Science, have not yet been thoroughly investigated. This paper fills the gap in the literature by systematically reviewing the usage of semantic technologies applied to Computer Science Curriculum Design.

\section{Research Methodology}
\label{sec:research_methodology}

\begin{figure}[h]
  \centering
  \includegraphics[width=\linewidth]{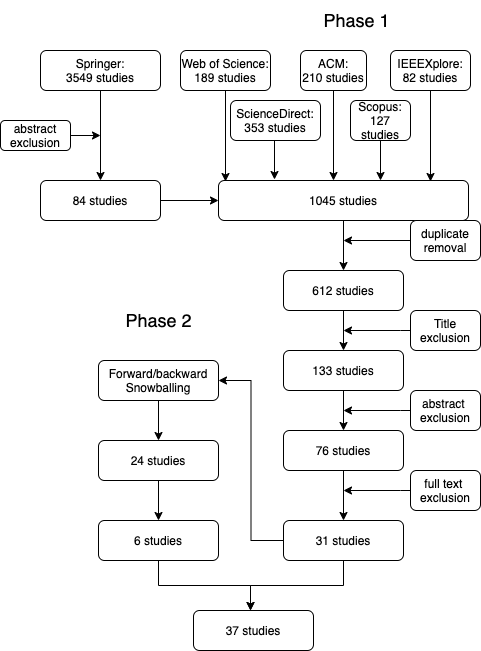}
  \caption{Paper Selection Process}
  \Description{papers}
  \label{img1}
\end{figure}

This systematic review was conducted following the methodology defined by Kitchenham and Charters \cite{Kitchenham2007}. The method is composed of three main steps: planning, conducting and reporting. The planning stage helps to identify existing research in the area of interest as well as build the research question. Our research question was created based on the PICOC method \cite{Kitchenham2007} and used to identify keywords and corresponding synonyms to build the search string to find relevant related works. The resulting search string is given below.

\emph{("computer science" OR "computer engineering" OR "informatics") AND ("curriculum" OR "course description" OR "learning outcomes" OR "curricula" OR "learning objects") AND ("semantic" OR "ontology" OR "linked data" OR "linked open data")}

To include a paper in this systematic review, we defined the following inclusion criteria: (1) papers must have been written in English; (2) papers must be exclusively related to semantic technologies, Computer Science and curriculum; (3) papers must have four or more pages (i.e., full research papers); (4) papers must be accessible online; and, finally, (5) papers must be scientifically sound, present a clear methodology and conduct a proper evaluation for the proposed method, tool or model.

Figure \ref{img1} shows the paper selection process in detail. Initially, 4,510 papers were retrieved using the ACM Digital Library\footnote{\url{https://dl.acm.org/}}, IEEE Xplore Digital Library\footnote{\url{https://ieeexplore.ieee.org/}}, Springer\footnote{\url{https://www.springer.com/}}, Scopus\footnote{\url{https://www.scopus.com/}}, ScienceDirect\footnote{\url{https://www.sciencedirect.com/}} and Web of Science\footnote{\url{https://www.webofscience.com/}} as digital libraries. The Springer digital library returned a total of 3,549 studies. The large number of papers returned by the Springer search mechanism led us to develop a simple crawling tool to help us in the process of including/rejecting papers in our systematic review. All the information returned by Springer was collected and stored in a relational database. After that, we were able to correctly query the Springer database and select the relevant papers for this systematic review. 

We also applied the forward and backward snowballing method to this systematic review to identify relevant papers that were not retrieved by our query string. The backward snowballing method was used to collect relevant papers from the references of the final list of papers in Phase 1, whereas the forward snowballing method was used to collect the papers that cited these papers \cite{Wohlin2014}. Google Scholar\footnote{\url{https://scholar.google.com/}} was used in the forward snowballing method. In total, 37 studies were identified as relevant; most of the studies were published in the last few years, although no relevant paper has been published in 2022. 

\section{Semantic Technologies in Computer Science Curriculum Design}
\label{sec:overview}

\subsection{Tools/Frameworks}
\label{subsec:tools}

Protégé\footnote{\url{https://protege.stanford.edu/}}\cite{Imran2016} is a popular open-source editor and framework for ontology construction, which several researchers have adopted to design curricula in Computer Science \cite{Tang2013,Wang2019,Nuntawong2017,Karunananda2012,Hedayati2016,Saquicela2018SimilarityDA,Maffei2016,Vaquero2009}. Despite its wide adoption, Protégé still presents limitations in terms of the manipulation of ontological knowledge \cite{Tang2013}. As an attempt to overcome this shortcoming, Asoke et al. \cite{Karunananda2012} developed a curriculum design plug-in called OntoCD, allowing curriculum designers to customise curricula by loading a skeleton curriculum and a benchmark domain ontology. OntoCD is evaluated by designing a Computer Science degree curriculum using a benchmark domain ontology proposed by  ACM and IEEE \cite{CS2013}. Moreover, Adelina and Jason \cite{Tang2013} used Protégé to develop the Sunway University Computing Ontology (SUCO), an ontology-specific Application Programming Interface for curricula management system. They claim that, in response to the shortcoming with using the Protégé platform, the SUCO tool shows a higher level of ability to manipulate and extract knowledge in addition to functioning effectively if the ontology is processed as an eXtensible Markup Language (XML) format document.

Other specific ontology-based tools in curriculum management have also been developed, e.g., the CDIO framework~\cite{Liang2012}. CDIO was created to automatically adapt a given curriculum according to teaching objectives and content. Similarly, Maffei et al. \cite{Maffei2016} utilised CONALI ontology as a tool for representing the semantics behind the constructive alignment activities to design, synthesise and evaluate courses in different domains. Likewise, in Mandić's study, the author presented a software platform\footnote{\url{http://www.pef.uns.ac.rs/InformaticsTeacherEducationCurriculum}} for comparing informatics teacher education curricula \cite{Mandic2018}. In Hedayati's work, the authors used the curriculum Capability Maturity Model (CMM), which is a taxonomical model used for describing the organisation’s level of capability in the domain of software engineering \cite{Paulk1993}, as the reference model to discuss an ontology-driven modelling to the culturally sensitive curriculum development process in the context of vocational ICT education in Afghanistan \cite{Hedayati2016}.

Finally, Vaquero et al. \cite{Vaquero2009} used an experience-sharing tool to transform information into knowledge to improve curriculum design called the ``Set of Experience Knowledge Structure'' (SOEKS) \cite{Sann2005SetOE}. SOEKS is used as part of a process to collect, infer and manage explicit knowledge using ontologies and semantic reasoning to provide curriculum designers with different expert perspectives for curriculum design.

\subsection{Datasets}
One of the most popular datasets used to build ontologies or as benchmarks between Computer Science curricula is the open-access CS2013\footnote{\url{https://cs2013.org/}} \cite{CS2013}. This dataset results from the joint development sponsored by the ACM and IEEE Computer Society of a computing curriculum \cite{Piedra2018}. CS2013 has been widely adopted \cite{Aeiad2016, Nuntawong2016, Nuntawong2017, Karunananda2012, Hedayati2016, Fiallos2018} due to its international reach and curricula and pedagogical guidelines.

Similar to CS2013, the Thailand Qualification Framework for Higher Education (TQF: HEd) was developed by the Office of the Thailand Higher Education Commission to be used by all Higher Education Institutions (HEIs) in Thailand as a framework to enhance the quality of course curricula and enable academic mobility. The TQF: HEd guidelines have been used in terms of ontology development in several studies for Computer Science curriculum design \cite{Nuntawong2017,Nuntawong2016,Hao2008,Nuntawong2015}.

Other studies use self-created datasets~\cite{Wang2019,Maffei2016,Hedayati2016,Fiallos2018}. Specifically, in Wang's work, the Ontology System for the Computer Course Architecture (OSCCA) was proposed based on a dataset created using course catalogues from top universities in China and network education Web sites \cite{Wang2019}. In Maffei's study, the authors experiment and evaluate the proposal based on the Engineering program at KTH Royal Institute of Technology in Stockholm, Sweden \cite{Maffei2016}. In Gubervic et al.'s work, the dataset used for comparing courses comes from the Faculty of Electrical Engineering and Computing at the University of Zagreb in Croatia and all universities from United States of America \cite{Guberovic2018}. In Fiallos's study, not only did the authors adopt CS2013 for domain ontologies modelling, but also the core Computational Sciences courses from the Escuela Superior Politécnica del Litoral (ESPOL\footnote{\url{https://www.espol.edu.ec/}}) in Ecuador were collected to compare semantic similarity between curricula. \cite{Fiallos2018}.

\subsection{Knowledge Representation}
RDF is used as the design standard for data interchange in the following studies \cite{Piedra2018,Nuntawong2017,Saquicela2018SimilarityDA}. In particular, Saquicela et al. \cite{Saquicela2018SimilarityDA} generated curriculum data in RDF format, creating and storing data in a repository when the ontological model has been defined and created.

Built on RDF and extended with additional vocabulary and semantics \cite{McGuinness2004}, OWL\footnote{\url{https://www.w3.org/OWL/}} is a vocabulary used in many studies for representing and sharing knowledge on the Web \cite{Piedra2018,Adrian2020,Mandic2018,Wang2019,Maffei2016,Vaquero2009}. Apart from OWL, two studies used XML\footnote{\url{https://www.w3.org/standards/xml/}} due to implementation requirements of their researches \cite{Tang2013,Hao2008}.

Body of Knowledge (BoK) is another method for representing knowledge, often using OWL. It offers a complete set of concepts, terms and activities for a professional domain \cite{Piedra2018}. BoK has been used in many studies \cite{Nuntawong2016,Hao2008,Karunananda2012,Tapia2018,Chung2014,Nuntawong2015} for curriculum design, including Piedra and Caro \cite{Piedra2018} who used it based on the IEEE/ACM CS2013 ontology as ``a specification of the content to be covered and a curriculum as an implementation''; Numtawong et al. \cite{Nuntawong2017} who used it to map different ontologies based on the ontology of TQF: HEd; and, Barb and Kilicay-Ergin \cite{Adrian2020} who used BoK based on the Library of Congress Subject Headings\footnote{\url{https://www.loc.gov/aba/cataloging/subject/}} to propose and evaluate an ontology for an Information Science curriculum.

\subsection{Use of Semantic Technologies in Computer Science Curriculum Design}
Semantic technologies have been found helpful in the context of curriculum design to (i) extract and analyse the relationship between concepts from learning materials; (ii) measure the similarity between courses/disciplines and their pre-requisites; and, (iii) create ontologies to enable course / curriculum analysis. These approaches mainly aim to enable academic mobility and curriculum interoperability across institutions \cite{Piedra2018}.

Most approaches and tools use natural language processing techniques (e.g., NLTK\footnote{\url{https://www.nltk.org/}}) to generate semantic descriptions for ontology creation \cite{Piedra2018,Tapia2018} based on extracted concepts and their relationships \cite{Wang2019}, find patterns in curricula \cite{Orellana2018} and compare them \cite{Kawintiranon2016}. Analysis and comparison of concepts extracted are performed based on word \cite{Orellana2018}, sentence \cite{Pawar2018,Aeiad2016,Fiallos2018} and concept network level \cite{Nuntawong2015,Nuntawong2016,Nuntawong2017,H.Gomaa2013}.

Lexical database of semantic relations such as WordNet\footnote{\url{https://wordnet.princeton.edu/}} \cite{Pawar2018} and Latent Semantic Analysis \cite{Deerwester90indexingby,Guberovic2018} are also used to measure the similarity of courses/disciplines or define ontology mapping rules to establish dependency relationships between curricula. For performance reasons, these approaches are often combined with clustering \cite{Saquicela2018SimilarityDA}, statistical methods \cite{Orellana2018,Adrian2020,Saquicela2018SimilarityDA}, and established taxonomies, such as the Bloom Taxonomy, \cite{Lasley2013,Mandic2018} to measure course similarity.

\section{Discussion}
\label{sec:challenges}
Semantic technologies have the potential to facilitate academic mobility, curriculum interoperability, program and course benchmarking, and the design of new curricula. However, despite its potential benefits, several challenges hamper its broad adoption in Computer Science and in other domains. 

One of the challenges is to create domain-specific ontologies to represent all the areas of computing, as mentioned in \cite{cc2020}. The main challenges are that the existing ontologies for CS represent different structures for CS programs and courses as well as represent concepts in different depths and contexts. Ontology Alignment (OA) techniques to address these issues have been extensively proposed \cite{Hao2008, Piedra2018}, however, OA is still a research problem \cite{DBLP:conf/dexa/NunesCCFLD13, DBLP:journals/ijwis/SouzaSN20}. Further investigation is required to understand the difference in CS curricula in different universities and how these differences can be used to enhance existing CS programs and curricula. Although Sekiya et al. \cite{Sekiya2015} present a study where the top ten universities in the USA in 2015 uniformly covered topics in Computer Science using the CS2013 curriculum guidelines, ``\emph{there is no single correct ontology for any domain}" \cite{Noy2001}. Therefore, multiple ontologies are required to account for the diversity and local needs of each CS program and course, HEI and country. We note, however, that reuse and modularity are important factors to be considered when building ontologies \cite{Rector2002} and, although a list of Good Ontologies\footnote{\url{https://www.w3.org/wiki/Good_Ontologies}} are available to assist ontology creators, many of the studies reviewed do not reuse them. 

Other aspects to consider when building ontologies are maintainability and correctness \cite{Rector2002}. An ontology must take into account changes and the correct interpretation of domain-specific concepts to avoid erroneous inferences and, therefore, impact the use of that ontology in a specific domain, e.g., in our context, hinder academic mobility and curricular interoperability. We recognise that this might be a difficult problem to solve as curriculum design are often based on cultural background and ideology.

Domain-specific semantic tools for curriculum design are lacking, while the use of multipurpose semantic tools can introduce errors into the process of creation (and alignment) of curriculum design, requiring manual verification \cite{Orellana2018}, which makes the process less attractive and, therefore, less adoptable by HEIs. To exemplify this problem, consider the common Learning Outcomes creation/alignment task. Many curriculum creators rely on the hierarchical structure of Bloom's Taxonomy (Remembering, Understanding, Applying, Analysing, Evaluating and Creating) \cite{Lasley2013}, but representing or capturing each of these levels from textual descriptions by multipurpose semantic tools is hard \cite{Mandic2018, Pawar2018}. Future research on semantic reasoners is promising, as seen in \cite{Vaquero2009}. 

For ontology creation, Protégé seems to be the standard tool (see Section \ref{subsec:tools}). However, as it is not a tool for creating curricula, many studies have created extensions (and other tools) to facilitate the creation of curricula \cite{Karunananda2012,Tang2013}. Future research on ontology creation tools and curriculum design is required to elicit the requirements for an ontology creation tool for curriculum design.

Another issue is related to the reproducibility of the proposed ontologies for CS curricula. Several studies \cite{Tang2013, Wang2019, Liang2012, Dexter2009, Vaquero2009} do not make their datasets available nor validate their approaches. We finalise this discussion with a call for reproducible research in this field. Ontologies and datasets must be available to take advantage of the benefits of semantic technologies.

\section{Conclusion}
\label{sec:conclusion}
This study presented a systematic review on semantic technologies applied in the context of Computer Science and Curriculum Design.

This paper addressed the most common topics covered in the literature, such as tools and frameworks used for curriculum design; datasets used to evaluate and propose Computer Science curricula; knowledge representation models to manage curriculum data; and the many uses of semantic technologies in the context of curriculum design.

Although there are common challenges faced in curriculum design using semantic technologies such as the efficiency of data processing and the evaluation and alignment of ontologies, the use of semantic technologies facilitates numerous improvements in academic activities (such as curricula design, mobility, interoperability and benchmarking). The literature reviewed are an initial but important step to allow the construction of CS curricula. However, to enable further adoption of semantic technologies, HEIs need to make their curricula available using semantic standards using frameworks and languages such as RDF and OWL.

\bibliographystyle{ACM-Reference-Format}
\bibliography{sample-base}

\appendix

\end{document}